\documentclass[12pt]{article}
\usepackage{amsmath}
\usepackage{graphicx}
\begin{document}

\title{Forward-backward multiplicity fluctuations in heavy nuclei collisions
in the wounded nucleon model}

\author{A. Bzdak$^{a,b}$ and K. Wozniak$^{a}$\\
$^{a}$ Institute of Nuclear Physics, Polish Academy of Sciences\\
Radzikowskiego 152, 31-342 Krakow, Poland\thanks{
e-mails: Adam.Bzdak@ifj.edu.pl, Krzysztof.Wozniak@ifj.edu.pl}\\
$^{b}$ Lawrence Berkeley National Laboratory, 1 Cyclotron Road\\ 
MS70R0319, Berkeley, CA 94720, USA }

\maketitle

\begin{abstract}
We use the wounded nucleon model to study the forward-backward multiplicity
fluctuations measured by the PHOBOS Collaboration in $Au+Au$ collisions at $%
\sqrt{s_{_{NN}}}=200$ GeV. The enhancement of forward-backward fluctuations
in $Au+Au$ collisions with respect to the elementary $p+p$ interactions is
in this model explained by the asymmetric shape of the pseudorapidity
density of produced particles from a single wounded nucleon and the
fluctuations of the number of wounded nucleons in the colliding nuclei. The
wounded nucleon model describes these experimental data better than the
HIJING, AMPT or UrQMD models do.

\vskip 0.6cm

\end{abstract}

\section{Introduction}

The studies of the fluctuations of the charged particles multiplicity in
heavy nuclei collisions provide information on the particle production and
the properties of the matter created in these collisions \cite{VK}. The multiplicity
measured in the symmetric pseudorapidity intervals in the forward and
backward hemisphere appears to be a suitable observable for such analysis.
Recently, the direct forward-backward correlations were studied by the STAR
Collaboration \cite{STAR-b}, while the difference between forward-backward
multiplicities was used by the PHOBOS Collaboration \cite{PHO-var}.

The PHOBOS Collaboration measured the variance $\sigma_{C}^{2}$ of the
forward-backward asymmetry variable $C=(N_{F}-N_{B})/\sqrt{N_{F}+N_{B}}$%
\begin{equation}
\sigma _{C}^{2}=\left\langle \frac{(N_{B}-N_{F})^{2}}{N_{B}+N_{F}}%
\right\rangle ,  \label{s2_def}
\end{equation}%
where $N_{B}$ and $N_{F}$ are the event-by-event multiplicities in the
backward $B$ and forward $F$ pseudorapidity intervals, respectively. The
measurement was performed for many different symmetric (with respect to $%
\eta =0$) pseudorapidity bins in the range $|\eta |<3$. It was found that:

(i) for $\eta $ bins with fixed width but changing position in
pseudorapidity $\sigma _{C}^{2}$ increases with increasing distance between
bin centers

(ii) for $\eta $ bins centered at the same position ($\eta =\pm 2$) $\sigma
_{C}^{2}$ increases with increasing width of the bins

(iii) $\sigma _{C}^{2}$ is significantly larger for peripheral (centrality $%
40-60\%$) collisions than for central (centrality $0-20\%$) collisions.

In this paper we use the wounded nucleon model \cite{WNM} to calculate the
values of $\sigma _{C}^{2}$ for the the same pseudorapidity bins as in the
PHOBOS experiment. First we perform calculations for the simple case of
elementary $p+p$ interactions and then we extend them to $Au+Au$ collisions
and different centralities. We show that the wounded nucleon model
reproduces reasonably well the experimental data, especially the centrality
dependence of the forward-backward fluctuations.

In the next section we define the model in detail. In Section 3 we describe
our Monte Carlo calculations and present the approximate analytical formula
for $\sigma _{C}^{2}$ derived in the framework of the wounded nucleon model
in the Appendix. In Section 4 our results are compared with the PHOBOS data
and also predictions of other models of nucleus-nucleus collisions are
discussed. We end our paper with conclusions.

\begin{figure}[th]
\begin{center}
\includegraphics[width=9cm]{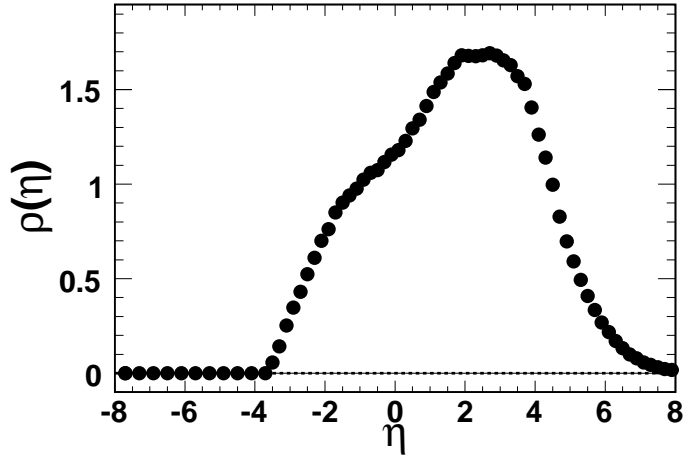}
\end{center}
\caption{The wounded nucleon fragmentations function $\protect\rho (\protect%
\eta )$: pseudorapidity density of produced particles from the
forward-moving wounded nucleon at $\protect\sqrt{s}=200$ GeV c.m. energy.}
\label{fig_wn}
\end{figure}

\section{Wounded nucleon model}

The wounded nucleon model \cite{WNM,ff-bc} assumes that after the collision
of two nuclei, the secondary particles are produced in the process of
independent fragmentation\footnote{%
It is similar to the assumption of independent hadronization of strings in
the dual parton model \cite{DPM}.} of these nucleons (called ''wounded
nucleons''), which underwent at least one inelastic collision. The
contribution from a wounded nucleon does not depend on the number of such
collisions\footnote{%
In fact in our calculations we will modify this assumption. We will come
back to this point later.}, thus the secondary particles are produced
according to a universal fragmentation function, $\rho (\eta )$. Number of
produced particles may vary following a multiplicity distribution.

The shape of the fragmentation function was extracted \cite{ff-bc,ff-bb}
from the PHOBOS data on $d+Au$ collisions \cite{PHO-dAu} at $\sqrt{s}=200$
GeV and is presented in Fig.~\ref{fig_wn} for a wounded nucleon from the
forward moving nuclei\footnote{%
An analogous wounded nucleon fragmentation function was constructed in Ref. %
\cite{ff-Ryb}.}. The fragmentation function for nucleons from the opposite
nuclei is a reflection with respect to $\eta =0$.

In this model the single particle density in elementary $p+p$ collisions is
described as $\rho (\eta )+\rho (-\eta )$. In the case of nucleus-nucleus
collision it becomes a sum of these functions multiplied by the number of
wounded nucleons in the appropriate nuclei \cite{ff-bc}: $\left\langle
w_{F}\right\rangle \rho (\eta )+\left\langle w_{B}\right\rangle \rho (-\eta
) $, where $\left\langle w_{F}\right\rangle $ and $\left\langle
w_{B}\right\rangle $ are the mean values of wounded nucleons in the forward
and backward moving nucleus, respectively. In consequence at midrapidity the
particle density in $A+A$ equals $\left\langle w\right\rangle \rho
(0)$ (where $\left\langle w\right\rangle =\left\langle w_{F}\right\rangle
+\left\langle w_{B}\right\rangle $) and is thus $\left\langle w\right\rangle
/2$ times larger than in elementary collisions. This prediction does not
agree with the experimental data, however, as the particle density $dN/d\eta
|_{\eta =0}$ is larger than $\left\langle w\right\rangle \rho (0)$ \cite%
{dataAuAu}. This can be easily corrected by introduction of a centrality
dependent factor $\gamma $ which appropriately increases the multiplicity of
particles generated in the fragmentation of a wounded nucleon that underwent
many inelastic collisions\footnote{%
Contrary to the wounded nucleon model assumption multiplicity from a wounded
nucleon slightly depends on the number of collisions it underwent.}.

The necessity of the above correction results from the composite structure
of the nucleons. It is accounted for in the wounded quark-diquark model \cite%
{ff-bb}, where the fragmenting objects are quarks and diquarks, and which
reproduces the particle density measured in $A+A$ collisions. The
fragmentation of a wounded nucleon is equal in this approach to a sum of
fragmentation functions of the quark and diquark weighted by the probability
of ''wounding'' each of them. The scaling factor $\gamma $ introduced by us
reflects the changes of these probabilities for different centralities of
the collisions. The wounded nucleon model with our correction is equivalent
to the wounded quark-diquark model in the kinematical area, where the
fragmentation of spectators (unwounded constituents) is negligible \cite%
{ff-bb}. This is fully justified in our case, as we are analyzing data
restricted to $|\eta |<3$.

The fragmentation function $\rho (\eta )$ represents the mean yield from a
wounded nucleon, whereas for calculation of fluctuations the changes
occurring event-by-event have to be included. We assume that the shape of $%
\rho (\eta )$ is unchanged, even if the integral of $\rho (\eta )$ follows a
multiplicity distribution. Further we also assume that the multiplicity
distributions in $p+p$ collisions in the combined pseudorapidity bin $B+F$
can be parameterized by negative binomial (NB) distribution \cite%
{UA5-200-n-k} 
\begin{equation}
P(n,\bar{n},k)=\frac{\Gamma (n+k)}{\Gamma (n+1)\Gamma (k)}\left( \frac{\bar{n%
}}{k}\right) ^{n}\left( 1+\frac{\bar{n}}{k}\right) ^{-n-k},  \label{nbd}
\end{equation}%
where $\bar{n}$ is the average multiplicity in the combined interval $B+F$, $%
1/k$ measures deviation from Poisson distribution and $\Gamma (n)$ is the
gamma function. In our calculations we need to know the multiplicity
distribution from a single wounded nucleon. Assuming that soft particle
production in $p+p$ collisions can be described by independent contributions
from each nucleon \cite{bz-pp}, it is straightforward to show that the
multiplicity distribution from a single wounded nucleon follows also NB
distribution (\ref{nbd}) but with $\bar{n}$ and $k$ substituted by $\bar{n}%
/2 $ and $k/2$, respectively\footnote{%
In general the multiplicity distribution from $w$ independent wounded
nucleons is given by NB distribution (\ref{nbd}) with $\bar{n}\rightarrow w%
\bar{n}/2$ and $k\rightarrow wk/2$, where $\bar{n}$ and $k$ are measured in $%
p+p$ collisions.}.

To perform the necessary calculations we also need to know the probability
that a particle originating from the forward moving wounded nucleon goes to $%
F$ interval, under the condition it is found either in $B$ or $F$. It can be
easily calculated as 
\begin{equation}
p=\frac{\int_{F}\rho \left( \eta \right) d\eta }{\int_{B}\rho \left( \eta
\right) d\eta +\int_{F}\rho \left( \eta \right) d\eta },  \label{p_def}
\end{equation}%
where $\rho \left( \eta \right) $ is the pseudorapidity density of produced
particles shown in Fig. \ref{fig_wn}. In consequence the probability that a
particle originating from the forward moving wounded nucleon goes to $B$
interval equals $1-p$. For the backward moving wounded nucleon $p$ has the
same meaning when the intervals $F$ and $B$ are exchanged. The $\bar{n}$
parameter from NB distribution can be also calculated using the
fragmentation function 
\begin{equation}
\bar{n}=2\int_{F}\rho \left( \eta \right) d\eta +2\int_{B}\rho \left( \eta
\right) d\eta .  \label{mn}
\end{equation}

For the nucleus-nucleus collisions in our modified version of the wounded
nucleon model the multiplicity of particles from a single wounded nucleon is
rescaled by $\gamma $ factor. It reflects the fact that a wounded nucleon
which underwent many inelastic collisions produces more particles than a
wounded proton in $p+p$ collisions. Accordingly, in Eq. (\ref{nbd})
we substitute $\bar{n}$ not by $\bar{n}/2$ but by $\gamma \bar{n}/2$. Less
obvious is the necessity of modification of $k$ parameter, which we decided
to multiply also by the factor $\gamma $. This is justified if we assume
that each nucleon is composed of two constituents, a quark and a diquark,
which populate particles independently. However, we show also results for
unmodified $k$ and we consider the difference between them as an additional
systematic uncertainty of the wounded nucleon model.

\section{Calculations}

In the present section we describe the MC calculation of $\sigma _{C}^{2}$
defined in Eq. (\ref{s2_def}) in the modified wounded nucleon model. In
addition we present also approximate analytical formulae (with $\bar{n}$ 
and $k$ rescaled be the factor $\gamma$) derived in the Appendix. 
We performed our calculations in three steps described below.

(A) In order to study effects of simple superposition of nucleon-nucleon
collisions we assume that in both nuclei we have fixed number of wounded
nucleons $w_{F}=w_{B}$. Each wounded nucleon adds it's contribution
generated independently according to the NB distribution (Eq. \ref{nbd})
with parameters $\bar{n}$ and $k$ substituted by $\gamma \bar{n}/2$ and $%
\gamma k/2$ respectively, where $\bar{n}$ and $k$ are measured in $p+p$
collisions. The generated particles are then assigned randomly, with
probability $p$ and $1-p$, to the intervals $F$ and $B$ (for the forward
moving wounded nucleon; to $B$ and $F$ for the backward moving nucleon).

Our main observation is that this contribution very quickly saturates with
increasing $w_{F}$. For $w_{F}>10$ it practically does not depend on $w_{F}$%
. In the Appendix we present the approximate analytical derivation of the
asymptotic formula for large $w_{F}$, which is: 
\begin{equation}
\left. \sigma _{C}^{2}\right| _{\text{A}}\approx 1+\frac{\bar{n}}{k}%
(2p-1)^{2}.  \label{s2A}
\end{equation}

This term depends only on the ratio $\bar{n}/k$, thus it is insensitive to
our modification of the wounded nucleon model, i.e. the simultaneous
substitution $\bar{n}\mapsto \gamma \bar{n}$ and $k\mapsto \gamma k$. Except
the most peripheral events $\left. \sigma _{C}^{2}\right| _{\text{A}}$ is
centrality independent.

(B) Contribution solely from the fluctuations in the number of wounded
nucleons can be obtained when we assume that each wounded nucleon fragments
always into the same, mean number of particles. Then each forward moving
wounded nucleon populates $\gamma \bar{n}p/2$ particles into the forward $F$
interval and $\gamma \bar{n}(1-p)/2$ particles into the backward $B$
interval (and analogously for the backward moving nucleons). In consequence
we obtain 
\begin{align}
N_{F}& =\frac{1}{2}\gamma \bar{n}\left[ w_{F}p+w_{B}(1-p)\right] ,  \notag \\
N_{B}& =\frac{1}{2}\gamma \bar{n}\left[ w_{B}p+w_{F}(1-p)\right] ,
\label{t1}
\end{align}%
where $w_{F}$ and $w_{B}$ are the numbers of wounded nucleons in the forward
and backward moving nuclei, respectively. The above formulae allow to
substitute $N_{F}$ and $N_{B}$ in the definition of $\sigma _{C}^{2}$ (Eq.~%
\ref{s2_def}): 
\begin{equation}
\left. \sigma _{C}^{2}\right| _{\text{B}}=\frac{1}{2}\gamma \bar{n}%
(2p-1)^{2}\left\langle \frac{(w_{F}-w_{B})^{2}}{w_{F}+w_{B}}\right\rangle .
\label{s2B}
\end{equation}

This term explicitly depends on $w_{F}$ and $w_{B}$ and thus is sensitive to
the centrality of the events sample. Moreover this term is sensitive to the $%
\gamma$ factor which was introduced to correct the wounded nucleon model.
The distribution of $w_{F}$ and $w_{B}$ cannot be obtained in a model
independent way. In order to match the experimental centrality definition we
are using the events from the HIJING generator, for which cuts compatible with 
these for the PHOBOS experimental data were performed \cite{phobos_priv}.
However, similar results can be obtained if $w_{F}$ and $w_{B}$ are obtained
from the Glauber type MC generator using centrality cuts on impact parameter
or on $w_{F}+w_{B}$\footnote{%
We checked it with our own MC generator with the Gaussian nucleon-nucleon
interaction profile.}.

(C) The full calculations combine (A) and (B) together and include both the
multiplicity distribution and the fluctuations of the number of wounded
nucleons. The $w_{F}$ and $w_{B}$ are taken from generated HIJING events.
Next each wounded nucleon populates particles according to the distribution (%
\ref{nbd}) in a procedure similar to that from (A).

Results of full simulations are practically identical with the sum of two
previously described terms (A) and (B). This is confirmed by the analytical
calculations in the Appendix as in the first approximation (neglecting terms
of the order $1/\left\langle w_{F}\right\rangle $) we obtain 
\begin{eqnarray}
\sigma _{C}^{2} &\approx &\left. \sigma _{C}^{2}\right| _{\text{A}}+\left.
\sigma _{C}^{2}\right| _{\text{B}}  \notag \\
&\approx &1+\bar{n}(2p-1)^{2}\left[ \frac{1}{k}+\frac{1}{2}\gamma
\left\langle \frac{(w_{F}-w_{B})^{2}}{w_{F}+w_{B}}\right\rangle \right] .
\label{s2all}
\end{eqnarray}

According to this formula, the values of $\sigma _{C}^{2}>1$ can be obtained
in the wounded nucleon model only when the wounded nucleon fragmentation
function is asymmetric with respect to $\eta =0$ (leading to $p\neq 0.5$).
The effects increasing $\sigma _{C}^{2}$, already contained in the
multiplicity distribution measured in elementary $p+p$ collisions are 
enhanced by superposition of nucleon-nucleon collisions. The fluctuations 
of the number of wounded nucleons, that cause the difference $w_{B}-w_{F}$ 
to be non-zero, further increase the $\sigma_{C}^{2}$.

This closes theoretical discussion of the problem.

\section{Results and discussion}

\label{sec_results}

In the present section we compare results for the variance $\sigma _{C}^{2}$
of the forward-backward asymmetry variable $C=(N_{F}-N_{B})/\sqrt{N_{F}+N_{B}%
}$ obtained in our wounded nucleon model MC simulations with the recently
published PHOBOS data \cite{PHO-var}. The measurement was performed for $%
Au+Au $ collisions at $\sqrt{s}=200$ GeV in the pseudorapidity range of $%
\left| \eta \right| <3$ for various symmetric forward and backward intervals.

\begin{table}[t!]
\begin{center}
\begin{tabular}{|c|c|c|c|}
\hline
$\eta $ & $\bar{n}$ & $k$ & $p$ \\ \hline
$0.25$ & $2.35$ & $2.0$ & $0.52$ \\ \hline
$0.75$ & $2.41$ & $2.3$ & $0.57$ \\ \hline
$1.25$ & $2.47$ & $2.6$ & $0.62$ \\ \hline
$1.75$ & $2.47$ & $2.9$ & $0.67$ \\ \hline
$2.25$ & $2.31$ & $3.2$ & $0.73$ \\ \hline
$2.75$ & $5.10$ & $3.5$ & $0.8$ \\ \hline
\end{tabular}
\hspace{1cm} 
\begin{tabular}{|c|c|c|c|}
\hline
$\Delta \eta $ & $\bar{n}$ & $k$ & $p$ \\ \hline
$0.25$ & $1.21$ & $3.0$ & $0.7$ \\ \hline
$0.50$ & $2.41$ & $3.0$ & $0.7$ \\ \hline
$1.00$ & $4.78$ & $3.0$ & $0.7$ \\ \hline
$1.50$ & $7.10$ & $3.0$ & $0.7$ \\ \hline
$2.00$ & $9.35$ & $3.0$ & $0.7$ \\ \hline
\end{tabular}%
\end{center}
\caption{The parameters used in the calculations of wounded nucleon model
predictions presented in Figs. \ref{fig_delta} and \ref{fig_eta}. We
estimate the uncertainty of the values of the parameters to be $7\%$, $20\%$
and $10\%$ for $\Delta \bar{n}/\bar{n}$, $\Delta k/k$ and $\Delta p/p$,
respectively. }
\label{tab}
\end{table}

\begin{figure}[ht]
\begin{center}
\includegraphics[width=14.5cm]{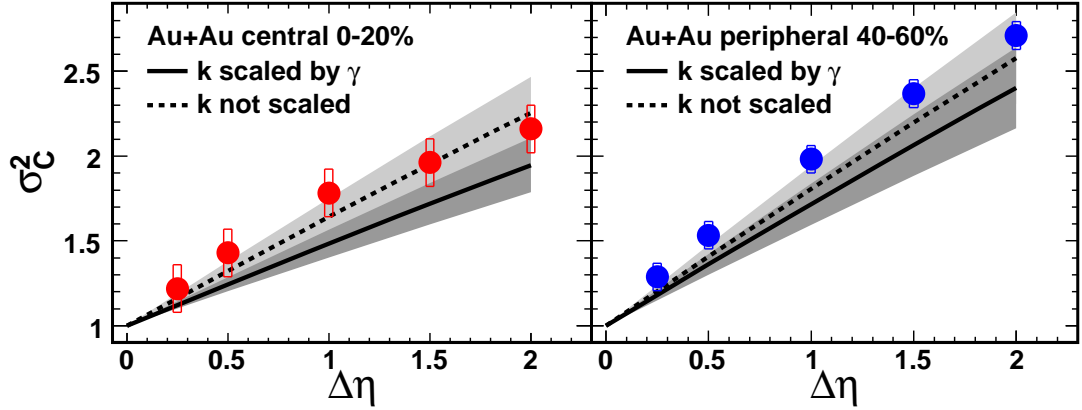}
\end{center}
\caption{The variance of the forward-backward asymmetry variable $%
C=(N_{F}-N_{B})/\protect\sqrt{N_{F}+N_{B}}$ as a function of the width of
the forward and backward intervals for most central collisions (left) and
the peripheral collisions (right). The position of the center of each bin is
fixed at $\protect\eta _{F}=2$ and $\protect\eta _{B}=-2$. The PHOBOS data
points (dots) are compared with the prediction of the modified wounded
nucleon model (continuous and dashed lines, for the scaled and unmodified $k$
parameter respectively). The grey bands reflect the systematic error of our
calculations, due to the uncertainty of the parameters (see the text near the end
of Section \ref{sec_results}). }
\label{fig_delta}
\end{figure}

\begin{figure}[ht]
\begin{center}
\includegraphics[width=14.5cm]{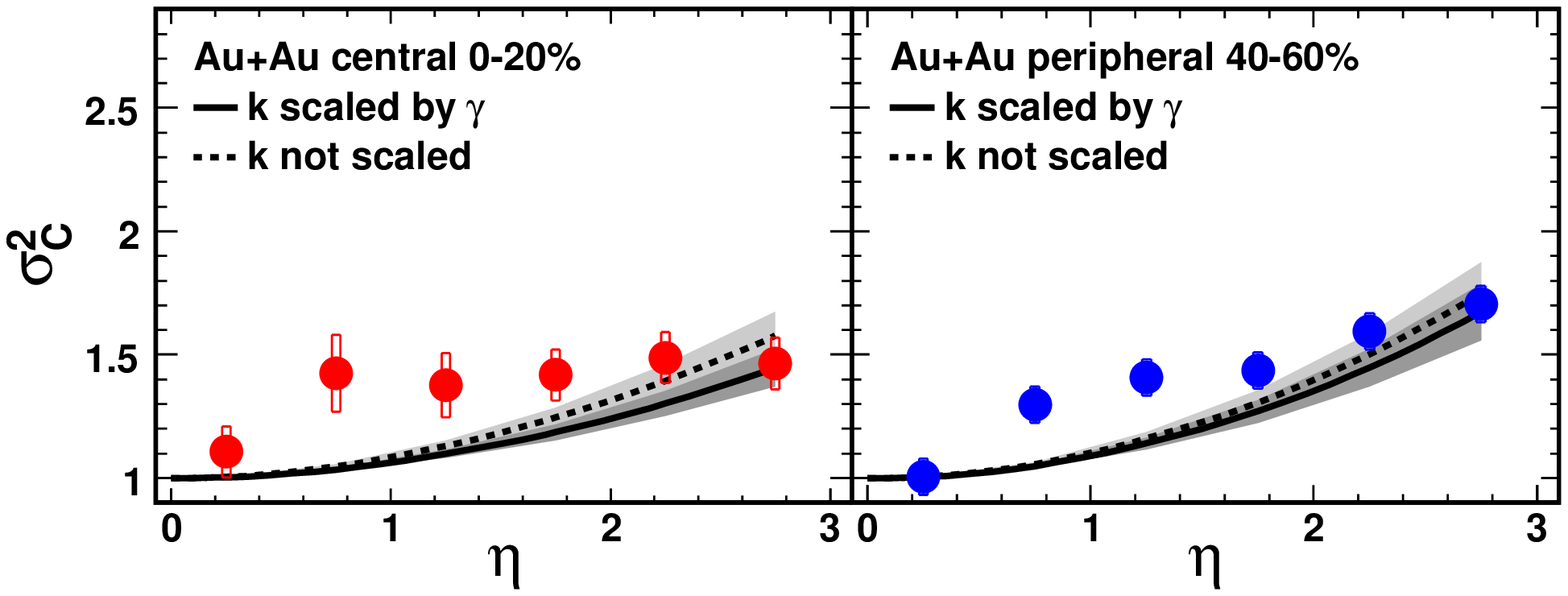}
\end{center}
\caption{The same as in Fig. \ref{fig_delta} but now the width of each bin
is fixed and equals $\Delta \protect\eta =0.5$, changing $\protect\eta $ is
the position of the center of the forward bin (the backward one is symmetric
around $\protect\eta =0$).}
\label{fig_eta}
\end{figure}

The model calculations were performed for the same intervals as in the
experiment and for the experimentally studied centrality classes. The
parameters $p$ and $\bar{n}$ were obtained from the wounded nucleon
fragmentation function shown in Fig. \ref{fig_wn} using Eqs. (\ref{p_def})
and (\ref{mn}) and are listed in Tab. \ref{tab}. The $\gamma $ factor
responsible for the correction of the wounded nucleon model was estimated by
comparing $\langle w\rangle \rho (0)$ and the PHOBOS data on $dN/d\eta $ at
midrapidity \cite{dataAuAu}, and equals $\gamma =1.35\pm 0.15$ and $1.6\pm
0.1$ for peripheral and central collisions, respectively.

The dependence of $\sigma _{C}^{2}$ on the bin width $\Delta \eta $
presented in Fig. \ref{fig_delta} includes points with the largest values of 
$\sigma _{C}^{2}$, measured with the smallest relative errors, and is thus
more challenging for the models. Here the center of each bin is fixed at $%
\eta _{F}=2$ and $\eta _{B}=-2$. Our modified wounded nucleon model
correctly reproduces the general trends observed in the data,
especially the centrality dependence. The predictions obtained for
the case with scaling of $k$ are slightly below the experimental
points, but in most cases are consistent within the systematic
errors. Interestingly, in the case of unmodified $k$, which we consider 
less probable, the agreement is much better.

In Fig. \ref{fig_eta} the values of $\sigma _{C}^{2}$ obtained for fixed 
width of each interval ($\Delta \eta =0.5$) and changing the distance between 
bin centers are shown. The wounded nucleon
model is in this case in worse agreement with the data, there are
discrepancies for $0.5<\eta <2$. However, for narrow $\eta $ bins the values
of $\sigma _{C}^{2}$ are closer to $1$, the systematic errors are relatively
larger and also the influence of effects not included in the model may be
thus stronger. The NB distribution which is supposed to contain effectively
all correlations present in the elementary collisions may be not sufficient
to fully represent the short range correlations (resonance decays, clusters).

\begin{figure}[th]
\begin{center}
\includegraphics[width=11cm]{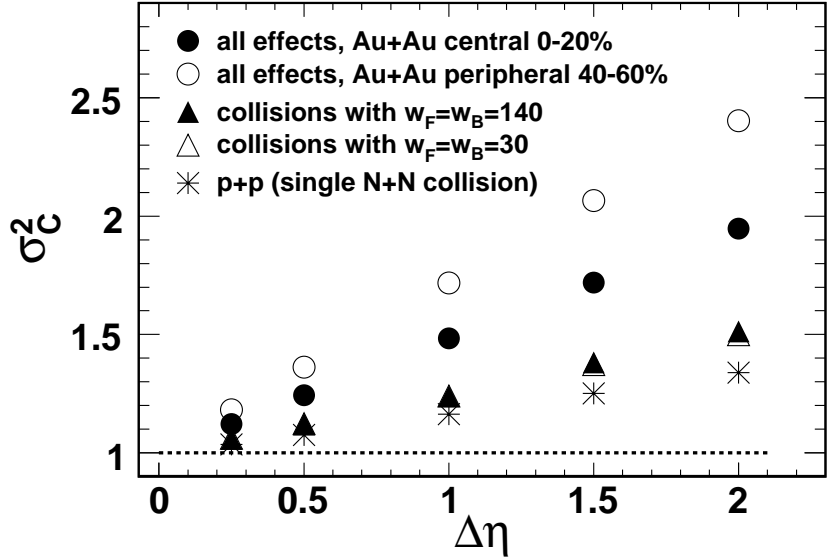}
\end{center}
\caption{Wounded nucleon model predictions (with scaled $k$ parameter) for $%
\protect\sigma _{C}^{2}$ values obtained for the pseudorapidity bins
centered at $\protect\eta =2$ with changing width $\Delta \protect\eta $.
The model calculations for central $0-20\%$ ($\langle w_{F}\rangle =140$)
and peripheral $40-60\%$ ($\langle w_{F}\rangle =30$) $Au+Au$ collisions
with all effects included (full and open circles) are compared with results
for collisions with exactly $140$ and $30$ wounded nucleons in each nuclei
(full and open triangles), the fluctuations in the number of wounded
nucleons is responsible for the difference between corresponding sets of
points. In addition the predictions for $p+p$ collisions are also shown
(stars). }
\label{fig_sklad}
\end{figure}

In Fig. \ref{fig_sklad} we show different components of $\sigma _{C}^{2}$
present in the wounded nucleons model (with scaled $k$ parameter), as
discussed in the previous section. For elementary $p+p$ collisions the value
of $\sigma _{C}^{2}$ for the widest interval ($\Delta \eta =2$) reaches $%
1.34 $, the superposition of nucleon-nucleon collisions increases it to
about $1.5 $. There is almost no difference between $w_{F}=30$ and $%
w_{F}=140$, which correspond to the mean values for the peripheral and
central collision analysed, as the superposition effects fast saturate. Much
stronger increase is due to contribution from the fluctuations of the number
of wounded nucleons, they add to the previous numbers directly and give $%
1.94 $ and $2.41 $ for central and peripheral collisions respectively.

Let us notice that all presented results\footnote{%
Except the predicted values of $\sigma _{C}^{2}$ for elementary $p+p$
collisions presented in Fig. \ref{fig_sklad}.} can be obtained from our
analytical approximation (\ref{s2all}), which agrees with the presented MC
results with the accuracy better that $3\%$. In the case of analytical
calculations we are using the values of $\left\langle \frac{%
(w_{F}-w_{B})^{2}}{w_{F}+w_{B}}\right\rangle $ extracted from samples of
peripheral and central HIJING events selected according to the experimental
centrality cuts, which are $0.38\pm 0.05$ and $0.90\pm 0.09$ for centrality $%
0-20\%$ and $40-60\%$, respectively \cite{phobos_priv}.

Finally, let us discus the uncertainty of our approach, represented by the
grey bands in Figs. \ref{fig_delta} and \ref{fig_eta}. The systematic errors
of model calculations can be estimated using the analytical formula (\ref%
{s2all}). We calculate the error of $\sigma _{C}^{2}$ caused by the
uncertainty of each parameter separately and add them in quadrature. In the
case of $k$ parameter the systematic error contains only the error of $k$ ($%
20\%$), the uncertainty of scaling $k$ for $A+A$ collisions is represented
by two extreme cases, for which the systematic errors are calculated
separately. All contributions due to errors of parameters are of the same
order, but the uncertainty of the probability $p$ has always the largest
impact. The systematic error of predicted $\sigma _{C}^{2}$ calculated this
way is approximately equal to $0.17$~($\sigma _{C}^{2}-1$).

\begin{figure}[th]
\begin{center}
\includegraphics[width=14.5cm]{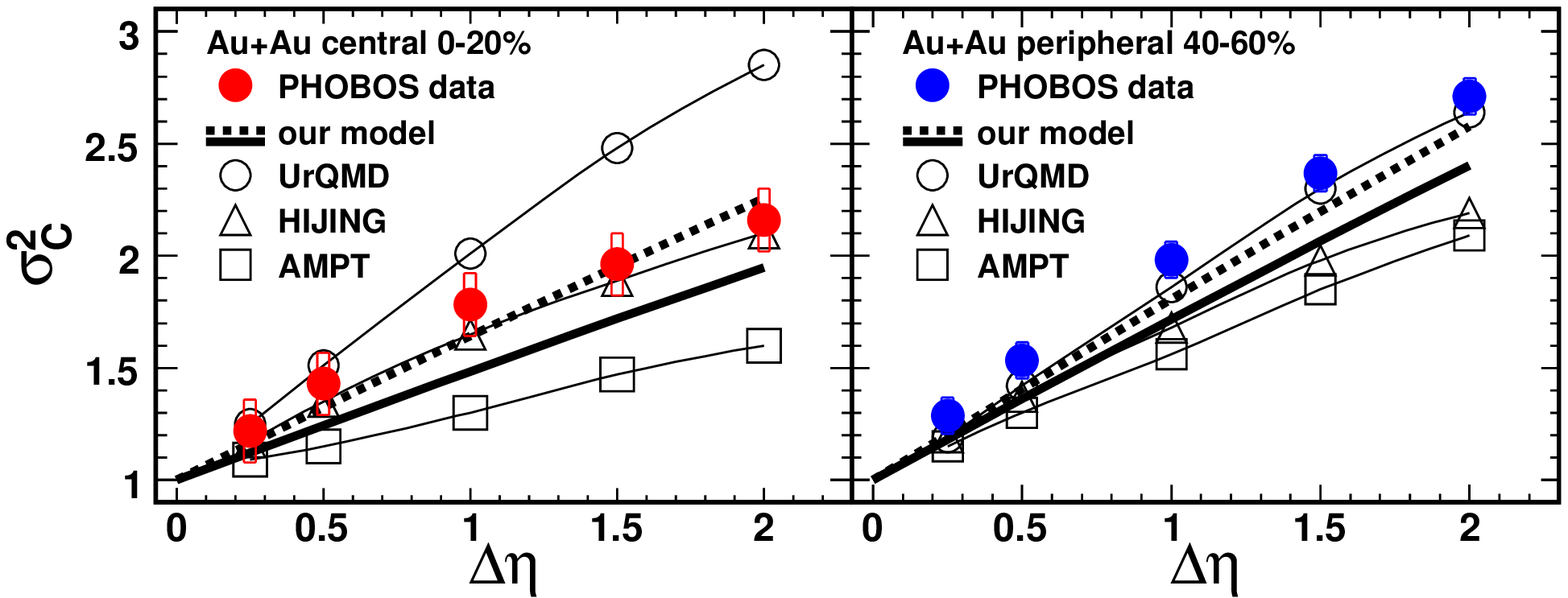}
\end{center}
\caption{Comparison of the experimental data with predictions from several
models for $\protect\sigma _{C}^{2}$ dependence on the width $\Delta \protect%
\eta $ of the pseudorapidity bin. Results for wounded nucleon model are
shown for two cases, $k$ parameter unscaled and scaled, as dashed and
continuous lines respectively. UrQMD \cite{urqmd}, HIJING and AMPT \cite{PHO-var}
predictions are represented by open symbols described in the plot. }
\label{fig_delta_models}
\end{figure}

The data analyzed in this paper were previously compared with expectations from
several models of nucleus-nucleus collisions: HIJING \cite{hijing}, AMPT %
\cite{ampt} in Ref. \cite{PHO-var} and UrQMD in Ref. \cite{urqmd}. As we can
see in Fig. \ref{fig_delta_models} the differences between them are large,
but none of these models correctly describes all features of the data listed
in the Introduction. HIJING gives approximately correct values of $\sigma
_{C}^{2}$ for central collisions, but they do not grow for peripheral
collisions. AMPT features expected centrality dependence, but always
significantly underestimates $\sigma _{C}^{2}$ values. UrQMD calculations
agree with the experimental results for peripheral events, but this model
predicts a slight increase of $\sigma _{C}^{2}$ for central collisions in
place of an observed significant drop. The predictions of the wounded
nucleon model are closest to the experimental data - especially for the
version with $k$ parameter not scaled.

\section{Conclusions}

Our conclusions can be formulated as follows.

(i) We show that the PHOBOS Collaboration data \cite{PHO-var} on the
variance $\sigma _{C}^{2}$ of the forward-backward asymmetry variable $%
C=(N_{F}-N_{B})/\sqrt{N_{F}+N_{B}}$ measured in $Au+Au$ collisions at $\sqrt{%
s}=200$ GeV can be reasonably described in the framework of the wounded
nucleon model. The wounded nucleon model predictions in most cases are
consistent within the errors with the data and slightly lower values of $%
\sigma _{C}^{2}$ may be due to short range correlations, which are only
partially accounted for in the model. Anyway, we reproduce both the
pseudorapidity and centrality dependence of $\sigma_{C}^{2}$ in contrast to
much more advanced HIJING, AMPT or UrQMD models.

(ii) The key ingredients of our approach are: the assumption of independent
hadronization of each wounded nucleon; the wounded nucleon multiplicity
distribution described by the negative binomial distribution (\ref{nbd})
and; the wounded nucleon pseudorapidity fragmentation function, shown in
Fig. \ref{fig_wn}. Moreover, we correct the wounded nucleon model by
taking into account the dependence of the multiplicity
from a single wounded nucleon on the number of collisions.

(iii) We also derive approximate analytical formula for $\sigma _{C}^{2}$ (%
\ref{s2all}). It can be observed that $\sigma _{C}^{2}-1$ is proportional to 
$(1-2p)^{2}$ where $p$ determines the partition of each particle between
forward and backward intervals. Thus the PHOBOS data can be reasonably
described only if the wounded nucleon fragmentation function is indeed
asymmetric in pseudorapidity, i.e., $p\neq 0.5$. It is also interesting to
note that the centrality dependence of $\sigma _{C}^{2}$ is fully determined
by the simple factor $\left\langle \frac{(w_{B}-w_{F})^{2}}{w_{B}+w_{F}}%
\right\rangle $.

\bigskip

\textbf{Acknowledgements}

We thank the PHOBOS Collaboration for the access to the simulation data
which enabled to calculate the model predictions most compatible with the
experimental conditions. Discussions with A. Bialas are highly appreciated.
This investigation was supported in part by the Polish Ministry of Science
and Higher Education, grant No. N202 125437, N202 034 32/0918, N202 282234 (2008-2010) and by the Director, Office of Energy Research, Office of High Energy and Nuclear Physics, Divisions of Nuclear Physics, of the U.S. Department of Energy under Contract No. DE-AC02-05CH11231. 
A.B. acknowledges support from the Foundation for Polish Science (KOLUMB programme).

\appendix

\section{Appendix: Calculation of $\protect\sigma _{C}^{2}$}

In this Appendix we present the approximate analytical calculation of $%
\sigma _{C}^{2}$. Let us introduce the following notation 
\begin{eqnarray}
N_{B} &=&\left\langle N_{B}\right\rangle (1+\varepsilon _{B}),  \notag \\
N_{F} &=&\left\langle N_{F}\right\rangle (1+\varepsilon _{F}).
\end{eqnarray}%
Although generally $\varepsilon _{B}$ and $\varepsilon _{F}$ may have any
value from $-1$ to infinity, if we exclude the most peripheral events, the
values of $|\varepsilon _{B}|$ and $|\varepsilon _{F}|$ are much smaller
than $1$. In such a case it is possible to expand the denominator of $\sigma
_{C}^{2}$ in the powers of $\varepsilon _{B}$ and $\varepsilon _{F}$ (here
we consider collisions of identical nucleus and symmetric $B$ and $F$ bins,
i.e. $\left\langle N_{B}\right\rangle =\left\langle N_{F}\right\rangle $) 
\begin{eqnarray}
\frac{1}{N_{B}+N_{F}} &=&\frac{1}{\left\langle N_{B}\right\rangle }\left( 
\frac{1}{2+(\varepsilon _{B}+\varepsilon _{F})}\right)  \notag \\
&=&\frac{1}{\left\langle N_{B}\right\rangle }\left( \frac{1}{2}-\frac{%
\varepsilon _{B}+\varepsilon _{F}}{4}+\frac{(\varepsilon _{B}+\varepsilon
_{F})^{2}}{8}+...\right) .  \label{expan}
\end{eqnarray}%
Taking the first term\footnote{%
Calculation of the higher orders is straightforward.} of this expansion into
account we obtain much simpler formula to handle%
\begin{equation}
\left\langle \frac{(N_{B}-N_{F})^{2}}{N_{B}+N_{F}}\right\rangle \approx 1+%
\frac{\left\langle N_{B}(N_{B}-1)\right\rangle -\left\langle
N_{B}N_{F}\right\rangle }{\left\langle N_{B}\right\rangle }.  \label{appr}
\end{equation}%
Indeed, now we can apply exactly the same method as in Ref. \cite{ab-star},
where the forward-backward correlation coefficient in the wounded nucleon
model was calculated. Let us construct the generating function 
\begin{equation}
H(z_{B},z_{F})=\sum%
\limits_{N_{B},N_{F}}P(N_{B},N_{F})z_{B}^{N_{B}}z_{F}^{N_{F}},  \label{H1}
\end{equation}%
where $P(N_{B},N_{F})$ is the probability to find $N_{B}$ particles in the
backward $B$ interval and $N_{F}$ particles in the forward $F$ one. Of
course $P(N_{B},N_{F})$ depends on the centrality of the collision. In Ref.~%
\cite{ab-star} we have shown that in the framework of the wounded nucleon
model, defined in Section 2, we obtain 
\begin{align}
H\left( z_{B},z_{F}\right) & =\sum\limits_{w_{B},w_{F}}W(w_{B},w_{F})\left\{
1+\frac{\bar{n}}{k}\left[ p\left( 1-z_{B}\right) +p^{\prime }\left(
1-z_{F}\right) \right] \right\} ^{-kw_{B}/2}\times  \notag \\
& \times \left\{ 1+\frac{\bar{n}}{k}\left[ p^{\prime }\left( 1-z_{B}\right)
+p\left( 1-z_{F}\right) \right] \right\} ^{-kw_{F}/2},  \label{H2}
\end{align}%
where $W\left( w_{B},w_{F}\right) $ is the probability distribution of the
numbers of wounded nucleons in the backward and forward moving nucleus,
respectively. $p^{\prime }=1-p$ ($B$ and $F$ are symmetric around $\eta=0$)
where $p$ is defined in Eq. (\ref{p_def}). $\bar{n}$ and $k$ come from the
NB fits (\ref{nbd}) to the $p+p$ multiplicity data in the combined bin $B+F$%
. Now the averages present in (\ref{appr}) can be directly calculated using
the generating function (\ref{H2}) 
\begin{align}
\left\langle N_{B}\right\rangle & =\left. \frac{\partial H(z_{B},z_{F})}{%
\partial z_{B}}\right| _{z_{B}=z_{F}=1},  \notag \\
\left\langle N_{B}(N_{B}-1)\right\rangle & =\left. \frac{\partial
^{2}H(z_{B},z_{F})}{\partial z_{B}^{2}}\right| _{z_{B}=z_{F}=1},  \notag \\
\left\langle N_{B}N_{F}\right\rangle & =\left. \frac{\partial
^{2}H(z_{B},z_{F})}{\partial z_{B}\partial z_{F}}\right| _{z_{B}=z_{F}=1}.
\label{poch}
\end{align}

Performing appropriate differentiations, multiplying $\bar{n}$ and $k$ by 
the $\gamma$ factor and taking the following relation into account
\begin{equation}
\sum\limits_{w_{B},w_{F}}W(w_{B},w_{F})\frac{(w_{B}-w_{F})^{2}}{%
2\left\langle w_{B}\right\rangle }=\frac{\left\langle
(w_{B}-w_{F})^{2}\right\rangle }{2\left\langle w_{B}\right\rangle }\approx
\left\langle \frac{(w_{B}-w_{F})^{2}}{w_{B}+w_{F}}\right\rangle ,
\end{equation}%
we obtain our final result (\ref{s2all}).


\begin{thebibliography}{99}

\bibitem{VK} See e.g., V. Koch, e-Print: arXiv:0810.2520 [nucl-th]. 

\bibitem{STAR-b} STAR Collaboration: B. I. Abelev et al., arXiv:0905.0237
[nucl-ex].

\bibitem{PHO-var} PHOBOS Collaboration: B.B. Back et al., Phys. Rev. C74
(2006) 011901.

\bibitem{WNM} A. Bialas, M. Bleszynski and W. Czyz, Nucl. Phys. B111 (1976)
461.

\bibitem{ff-bc} A. Bialas and W. Czyz, Acta Phys. Polon. B36 (2005) 905.

\bibitem{DPM} A. Capella, U. Sukhatme, C-I Tan and J. Tran Thanh Van, Phys.
Rept. 236 (1994) 225.

\bibitem{ff-bb} A. Bialas and A. Bzdak, Phys. Rev. C77 (2008) 034908; Phys.
Lett. B649 (2007) 263; Acta Phys. Polon. B38 (2007) 159. For a review, see
A. Bialas, J. Phys. G35 (2008) 044053.

\bibitem{PHO-dAu} PHOBOS Collaboration: B.B. Back et al., Phys. Rev. C72
(2005) 031901.

\bibitem{ff-Ryb} G. Barr, O. Chvala, H.G. Fischer, M. Kreps, M. Makariev, C.
Pattison, A. Rybicki, D. Varga and S. Wenig, Eur. Phys. J. C49 (2007) 919;
A. Rybicki, Acta Phys. Polon. B33 (2002) 1483.

\bibitem{dataAuAu} See e.g., PHOBOS Collaboration: B.B. Back et al., Phys.
Rev. C65 (2002) 061901.

\bibitem{UA5-200-n-k} UA5 Collaboration: R.E. Ansorge et al., Z. Phys. C43
(1989) 357.

\bibitem{bz-pp} A. Bzdak, e-Print: arXiv:0906.2858 [hep-ph].

\bibitem{phobos_priv} PHOBOS Collaboration (MC simulations), private
communication.

\bibitem{urqmd} S. Haussler, M. Abdel-Aziz and M. Bleicher, Nucl. Phys. A785
(2007) 253.

\bibitem{hijing} X.N. Wang and M. Gyulassy, Phys. Rev. D44 (1991) 3501.

\bibitem{ampt} Z.W. Lin, C.M. Ko, B.A. Li, B. Zhang and S. Pal, Phys. Rev.
C72 (2005) 064901.

\bibitem{ab-star} A. Bzdak, Phys. Rev. C80 (2009) 024906.
\end{thebibliography}
\end{document}